# Learning in the Large -
## *An Exploratory Study of Retrospectives in Large-Scale Agile Development*

Torgeir Dingsøyr[1,2], Marius Mikalsen[1], Anniken Solem[1], Kathrine Vestues[2]

[1]SINTEF, 7465 Trondheim, Norway
{torgeir.dingsoyr|marius.mikaelsen|anniken.solem}@sintef.no
[2]Department of Computer and Information Science,
Norwegian University of Science and Technology
kathrine.vestues@ntnu.no

**Abstract.** Many see retrospectives as the most important practice of agile software development. Previous studies of retrospectives have focused on process and outcome at team level. In this article, we study how a large-scale agile development project uses retrospectives through an analysis of retrospective reports identifying a total of 109 issues and 36 action items as a part of a longitudinal case study. We find that most of the issues identified relate to team-level learning and improvement, and discuss these findings in relation to current advice to improve learning outcome in large-scale agile development[1].

## 1 Introduction

Retrospective meetings are extremely important in agile methods. The Agile Practice Guide describes them as "*the single most important practice in agile development*" [1] and in his much read book on Scrum and XP, Kniberg states that retrospectives are the "*number-one-most-important thing in Scrum*" [2]. According to the 11th State of Agile report [3], the retrospective is the third most used agile practice. We find many suggestions on how to conduct retrospectives in the agile practitioner literature such as [4, 5] and online[2].

The purpose of retrospectives is to explore the work results of a team in an iteration or a phase in order to "*learn about, improve, and adapt its process*" [1]. The advice offered in the agile community has mainly focused on learning and improvement for the team, while such practices also have a potential to provide learning both on the individual level and for a larger project or organization.

In this article, we focus on the practice of learning and improving through retrospectives in large-scale agile development. The research agenda for large-scale agile development has identified knowledge-sharing as an important topic [6]. This is a particularly challenging area of work, as such projects consists of several development teams with dependencies between teams and typically involve complex integration with existing ICT systems in projects that are critical for companies or societies [7].

We structure this article as follows: First, we provide background on studies of retrospective practices and prior studies on analysing content and effect of retrospectives and formulate research questions. Second, we present an exploratory case study and method for analysis of retrospectives. Third, we discuss what the

---

retrospectives have addressed and what could be done to improve the learning outcome of retrospectives in the large and suggest directions for future research.

## 2   Background

Given the importance of retrospectives in agile development, the topic has received relatively little attention in scientific studies. A review of previous studies on IT retrospectives finds a multitude of definitions of retrospectives, descriptions of a number of outcomes, different practices described, and "*no project retrospective measurements given to confirm [...] whether outcomes have been successfully achieved*" [8].

Kniberg [2] describes a retrospective practice as a team exercise lasting 1-3 hours where a team identifies what has been «good», what «could have been better» and «improvements», and suggest voting on the improvements to focus on in the next iteration. The practices described in the research literature [8] typically involve additional steps, for example including root cause analysis in order to analyse topics identified before deciding on action items to include in the next iteration.

In a study of retrospective practices at team level, Lethinen et al. [9] found that most discussions were related to topics close to and controllable by the team, but that topics that could not be resolved at the team level due to their complexity nevertheless recurred over time.

Many previous studies have seen retrospectives as an arena for reflection to enable learning and process improvement [10]. Andryiani et al. [11] studied retrospectives with a framework describing stages of reflection as reporting, responding, relating, reasoning and reconstructing. A finding is that agile teams may not achieve all levels of reflection simply by performing retrospective meetings. The study found that "*important aspects discussed in retrospective meetings include identifying and discussing obstacles, discussing feelings, analysing previous action points, identifying background reasons, identifying future action points and generating a plan*" [11].

We have not been able to identify studies of retrospectives in large-scale development, but a blog post describes how Spotify conducted large-scale retrospectives[3] in «one big room» in a format similar to world café [12]. Advice in one of the large-scale development frameworks, Large-Scale Scrum (LeSS),[4] is to hold an «overall retrospective» after team retrospectives, to discuss cross-team and system-wide issues, and to create improvement experiments.

We do not know of studies investigating the learning effect of retrospectives, but a summary of relevant theories of learning such as Argyris and Schön's theory of learning and Wenger's communities of practice can be found in one overview article [13], which discusses learning on individual-, team-, and organizational level. Argyris and Schön distinguish between smaller improvement («single loop learning») and more thorough learning («double loop learning»).

In this explorative study, we ask the following research questions: *1. How are retrospectives used in a large-scale agile development project? 2. What could be done to improve the learning outcome of retrospectives in large-scale agile projects?*

## 3   Method

We are currently conducting a longitudinal case study [14] of a large-scale development project. The case was selected as it is one of the largest development projects in Scandinavia, and is operating in a complex environment with heavy integration with other ICT systems.

The customer organization has 19 000 employees, and close to 300 ICT systems. A new solution will require changes to 26 other systems. The project uses a stage-gate delivery model with 4 stages (analysis of needs, solution description, construction, and approval, similar to a previous project described in [15]). We followed the first release, with 37 developers in four development teams. Teams had a Scrum-master, one or two application architects, one or two testers, and up to ten developers. The project uses the Scrum-method, with three-week iterations, starting with a planning meeting and ending with a demo and retrospective.

---

[3] https://labs.spotify.com/2015/11/05/large-scale-retros/
[4] https://less.works/less/framework/index.html

The project has three main releases, and this article is based on an analysis of minutes of meetings from 10 retrospectives in the first release. The minutes include iterations 3 to 9, with an exception of iteration 6, when no retrospective was held due to summer holidays. The minutes cover a 5-month period.

We have limited the study to an analysis of retrospective minutes from two of the four teams. The minutes describe who were present in the face-to-face meeting, a list of issues that went well, a list of issues that could be improved and most often a list of action items. The length of the minutes varied from half a page to two pages. The minutes were posted in the project wiki.

We all read three minutes individually, and then jointly established a set of categories, taken from the Scrum guide[5], which describes the purpose of the sprint retrospective as an arena for inspecting how the last sprint went with regards to the categories «people», «relationships» (merged with people), «process», and «tools». We added the categories «project» and «other teams» to specifically address the large-scale level. These categories were used to code issues and action items.

## 4 Results

Bra / bedre uten tiltak

The analysis of minutes from retrospectives in Table 1 shows the issues recorded by the teams during the seven iterations. Most issues were related to «process« (41) and «people and relationships» (30). In the following, we describe issues that emerged in selected categories, and then present the resulting action items as recorded in the minutes.

**Table 1.** Issues that went well and issues that could be better. In total 109 issues were recorded during seven iterations for two teams. Roughly 40% of issues were statements on issues that went well and 60% about issues that could be improved.

|  | Iteration 3 | Iteration 4 | Iteration 5 | Iteration 6 | Iteration 8 | Iteration 9 | Iteration 10 | Sum |
|---|---|---|---|---|---|---|---|---|
| *Process* | 7 | 3 | 7 | 8 | 5 | 7 | 4 | 41 |
| *People & relationships* | 1 | 1 | 13 | 5 | 4 | 6 | 0 | 30 |
| *Other topics* | 1 | 2 | 3 | 0 | 2 | 2 | 2 | 12 |
| *Project* | 0 | 0 | 4 | 0 | 1 | 3 | 2 | 10 |
| *Tools* | 1 | 0 | 3 | 1 | 1 | 1 | 2 | 9 |
| *Other teams* | 2 | 0 | 1 | 2 | 2 | 0 | 0 | 7 |

Due to space limitations, the following results describe the issues we found relating to the categories that shed most light on how large-scale agile development influences the teams. These are «process» (41 reported issues), «project» (10 reported issues) and «other teams» (7 reported issues).

In terms of *process*, there were issues such as that the build breaks too often, design takes too much capacity from the team, that they would like more consistent use of branching in Git (tool for version control and code sharing), and that frequent check-ins makes it difficult to terminate feature-branches. The following excerpt illustrates how process issues manifest: "*A lot of red in Jenkins [a continuous integration tool], which makes it difficult to branch from «develop»*". Other issues were concerned with quality control and routines in the team, such as the need for better control and routines for branching of the code, need for more code reviews, too many and messy Jira (issue tracker) tasks, and architects have limited time to follow up on development. Issues concerning lack of structure for bug reporting were reported as such: "*Structure concerning tests, bugs are reported in all possible ways – Mails – Skype – face to face, very difficult to follow up and have continuity in test/bugfix etc.*"

*Project* issues are related to the overall organisation of the project as a whole. Such issues were far less frequently reported, and those we found included having updated requirements for user stories when entering sprints, that solutions designs should be more detailed, product backlog elements should be ready before sprint start, and addressing how developers move between teams. The following illustrates how one team reports the need for more technical meetings between teams on a project level: "*Review of code/project, all meetings are*

---
[5] http://www.scrumguides.org/

*about organisation, but it should be one meeting about how our code/setup/project looks from a technical perspective*".

Finally, for the category *other teams*, i.e. how teams interact in a multi-team setting, we found how there were issues with regard to how teams "takes instructions" from several different parties, and how there was challenges in detecting dependencies in the code before you develop and test. The following excerpt from the retrospective minutes illustrates how one team is not involved sufficiently in the planning of refactoring: "*We want to be notified in advance when there are big refactorings or other significant changes in the codebase, before it happens*".

The retrospective minutes also contains actions decided on by the teams. In total, the two teams identified 36 action items, where most were related to «process» and to «other topics». We show the distribution and provide examples of action items in Table 2.

**Table 2.** Action items from retrospective minutes according to topic.

| *Topic* | *Number* | *Example action items* |
| --- | --- | --- |
| Process | 13 | "Review and assign quality assurance tasks during daily stand-up." |
| Other topics | 7 | "We need a course on the «react» technology." |
| Tools | 5 | "More memory on the application development image." |
| People and relationships | 5 | "Organise an introduction round for new team members." |
| Project | 4 | "Have backlog items ready before an iteration starts." |
| Other teams | 2 | "Be more aware of dependencies when assigning tasks, make sure that other teams we depend on really give priority to these tasks." |

## 5. Discussion

We return to discuss our two research questions, starting with *how are retrospectives used in a large-scale agile development project?*

We found that retrospectives were used at team level, where short meetings were facilitated by the scrum master and reported in minutes on the project wiki. Minutes were available to everyone in the project, including customer representatives.

Our analysis of topics addressed in the retrospectives shows that most of the issues identified as either «working well» or «could be improved» related to *process*, followed by *people and relationships*. In the «large-scale» categories *project* and *other teams* we found in total 17 issues of the total 109. However, as shown in the results, many of the issues described as *process* were related to the scale of the project, such as identifying challenges with the merging of code or detailing of specifications before development would start. We find, however, that teams mainly deal with team-internal issues in retrospectives.

The analysis of the action items shows that 6 of the 36 action items identified during the work on the release were in the «large-scale» categories. However, we see that some of the action items in the other categories are related to scale. One example is the item "organizing an introduction round for new team members" in the category *people and relations*, which describes an action item which would not be necessary on a single-team project. However, our impression is also here that most action items concern issues at the team level.

We have not been able to conduct an analysis of the effect of retrospectives at team level. We consider that such meetings give room to develop a common understanding of development process, tasks and what knowledge people in the team possess, what in organizational psychology is referred to as *shared mental models* [13] and have been shown to relate to team performance. A common critique of retrospectives is that teams meet and talk, but little of what is talked about is acted upon. We have not been able to assess how many of the 36 identified action items were acted upon, but found in one minute that "all action items suggested to the project management has been implemented". The 36 action items identified can be considered small improvement actions. Given the short time spent on retrospectives, they do not seem to facilitate «deep» learning («double loop» learning in Argyris and Schön's framework). Having minutes public could also lead to critique being toned down or removed completely.

This leads us to discussing our second research question - w*hat could be done to improve the learning outcome of retrospectives in large-scale agile projects?*

In the background we pinpointed particular challenges of large-scale agile development such as dealing with a high number of people and many dependencies [7]. A retrospective can be used for a number of purposes. Prior studies in organizational psychology suggest that in projects with many teams, the coordination between teams are more important than coordination within teams [16]. It is reason to believe it would be beneficial to focus attention on inter-team issues in large projects. The LeSS framework suggests organizing inter-team retrospectives directly after the team retrospectives. Alternatively, teams can be encouraged to particularly focus on inter-team issues as part of the team retrospectives. A challenge in the project studied is that the contract model used may hinder changes, for example the contract model specifies handover phases between companies involved in the *analysis of needs* phase and the *solution description* and *development phase*. However, given the limitations, it is important that the project adjusts work practice also on inter-team level to optimize use of limited resources.

This exploratory study has several limitations, where one is that we have only analysed minutes available on the project wiki from two of four teams.

# 6. Conclusion

Many in the agile community regard retrospectives as the single most important practice in agile development. It is therefore interesting to know more about how retrospectives are practiced in large-scale development where there is a dire need to learn and improve as many participants are new to the project, the customer organization, and to the development domain. We found that short retrospectives were conducted at team level and mostly addressed issues at the team-level. The action items mainly addressed team level issues. Most actions also seem to relate to smaller improvements, what Argyris and Schön call «single-loop learning».

A large-scale project will benefit from learning and improvement on the project level, and this would be strengthened by following the advice from LeSS by facilitating retrospectives at the project level. Further, to shift learning effects towards «double-loop learning», we suggest that more time is devoted to the retrospectives.

In the future, we would like to initiate retrospectives at the inter-team level, explore the types of issues that are raised, and also gain more knowledge about perceptions of retrospectives by interviewing project participants.

**Acknowledgement.** This work was conducted in the project Agile 2.0 supported by the Research Council of Norway through grant 236759 and by the companies Kantega, Kongsberg Defence & Aerospace, Sopra Steria, Statoil and Sticos.